\documentclass[12pt]{article}
\usepackage[dvipsnames]{xcolor}
\usepackage{pgfplots}
\pgfplotsset{width=7cm,compat=1.17}
\usetikzlibrary{patterns}

\newcommand{\be}{\begin{equation}}
\newcommand{\ee}{\end{equation}}
\newcommand{\bey}{\begin{eqnarray}}
\newcommand{\pslash}{\not{\hbox{\kern-2.3pt $p$}}}
\newcommand{\pdslash}{\not{\hbox{\kern-2pt $\partial$}}}

\newcommand{\beq}{{\begin{equation}}}
\newcommand{\eeq}{{\end{equation}}}
\newcommand{\bea}{{\begin{array}}}
\newcommand{\eea}{{\end{array}}}
\newcommand{\ri}{{\rm i}}

\newcommand{\eey}{\end{eqnarray}}

\begin{document}

\begin{titlepage}
\vskip 2cm
\begin{center}
{\Large\bf Non-relativistic Fermi particle in one-dimensional pseudoscalar $\delta$-function potential}
\vskip 3cm
{\bf
Fuad M. Saradzhev}
\footnote{{\tt fuads@athabascau.ca} }
\vskip 5pt
{\sl Centre for Science, Athabasca University,
 Athabasca, Alberta, Canada \\}
\vskip 2pt

\end{center}
\vskip .5cm
\rm

\begin{abstract}

It is shown that a non-relativistic Fermi particle with a non-zero rest energy moving in pseudoscalar ${\delta}$-function potential in one dimension can be confined for both signs of the coupling constant.
The binding energies depend on the value of the particle's rest energy, and in the limit of vanishing
rest energy only one of the bound states survives. The coefficients of reflection and transmission are
determined, and the conditions for complete reflection and transmission are discussed.

\end{abstract}

\end{titlepage}

\setcounter{footnote}{0} \setcounter{page}{1} \setcounter{section}{0} %
\setcounter{subsection}{0} \setcounter{subsubsection}{0}

\section{Introduction}

The dynamics of non-relativistic Fermi particles moving in pseudoscalar external potentials is affected
by their rest energies. This was shown in \cite{sar} for a pseudoscalar spherical well of finite depth and a pseudoscalar Coulomb potential. The Fermi particles with non-zero rest energies have a set of bound states which does not exist for the ones with vanishing rest energies.

The rest energy effects show themselves in the framework of the L\'{e}vy-Leblond (LL) equation \cite{levy} generalized to include a rest energy term. The generalized L\'{e}vy-Leblond (GLL) equation \cite{sar2011} can be reduced to a Schr\"odinger type equation with a rest energy dependent effective potential. This allows us to find the impact of rest energy on the bound state spectrum.

In this work, we study the motion of a non-relativistic Fermi particle in pseudoscalar $\delta$-function potential in one dimension. For $\delta$-function potentials in the one-dimensional Schr\"odinger equation, both the bound and scattering states are well known \cite{flugge} . A single bound state exists in the attractive $\delta$-function potential only. For scattering at $\delta$-function potentials, complete reflection is observed when the energy of the incident particle approaches zero, $E \to 0$, and complete transmission occurs when the energy is infinitely large, $E \to \infty$. We aim to determine if this picture is valid for pseudoscalar $\delta$-function potentials as well and how it changes if the rest energy is taken into account.

Our paper is organized as follows. In Sect. 2, we introduce the one-dimensional GLL equation with a pseudoscalar $\delta$-function potential and give its equivalent form with the rest energy dependent effective potential. We find the bound state spectrum in Sect. 3 and the scattering amplitudes in Sect. 4. We conclude with discussion in Sect. 5.

\section{GLL equation}

The GLL equation for a $4$-component non-relativistic Fermi field
${\psi}({\bf r},t)$ with inertial mass $m$ and rest energy $E_0$ reads \cite{sar2011}
\begin{equation}
\left({\rm i}{\hbar}
\gamma^{\bar{\mu}}\partial_{\bar{\mu}}-kI_{+} - V({\bf r},t) \right){\psi}({\bf r},t)=0,
\label{compact}
\end{equation}
where $\bar{\mu}$ runs from $1$ to $4$, ${\bf r}=(x^1=x, x^2=y, x^3=z)$, ${\Delta} \equiv {\partial}_a^2$, ${\partial}_a \equiv \frac{\partial}{{\partial}x^a}$, $a=1,2,3$, ${\partial}_4 \equiv \frac{1}{c} \frac{\partial}{{\partial}t}$ and
\[
I_{+} \equiv I - \frac{m{c}}{k} {\gamma}^5,
\]
$I$ being the identity matrix.

A $4 \times 4$ matrix-valued function $V({\bf r},t)$ represents an external potential, and $k$ is the momentum corresponding to the rest energy, $E_0 = k^2/(2m)$. The ${\gamma}$-matrices
\bey
\gamma^a=
\left(\begin{array}{cc}
0 & \ri\sigma^a\\
\ri\sigma^a & 0
\end{array}\right),\;\;\;
\gamma^{4}=\frac{1}{\sqrt{2}} \left(
\begin{array}{cc}
1 & 1 \\
-1 & -1
\end{array}\right),\;\;\;
\gamma^5=\frac{1}{\sqrt{2}} \left(
\begin{array}{cc}
1 & -1\\
1 & -1
\end{array}\right) \nonumber
\eey
fulfil the algebra
\[
\gamma^{\mu} \gamma^{\nu} + \gamma^{\nu} \gamma^{\mu} = 2g^{\mu \nu},
\]
with ${\mu},{\nu}=1,...,5$ and
\[
g^{\mu\nu}=\left( \begin{array}{ccc} -{\mathbf 1}_{3\times 3}
& 0 & 0 \\ 0 & 0 & 1 \\ 0 & 1 & 0\end{array} \right).
\]
For $k=0$, Eq.(\ref{compact}) reduces to the LL equation \cite{levy}:
\[
\left({\rm i}{\hbar}
\gamma^{\bar{\mu}}\partial_{\bar{\mu}} + mc {\gamma}^5 - V({\bf r},t) \right){\psi}({\bf r},t)=0.
\]

Let us assume that the field ${\psi}$ only depends on one spatial coordinate $z$, and let us consider the motion of this field in the time-independent pseudoscalar potential
\begin{equation}
V = - \frac{1}{c} g{\gamma}^5 {\delta(z)},
\label{potone}
\end{equation}
where $g$ is a coupling constant. Then the one-dimensional version of the GLL equation is
\begin{equation}
\left(\frac{\rm i}{c} {\gamma}^4 \frac{\partial}{{\partial}t} +
{\rm i} \gamma^{3} \frac{\partial}{{\partial}z} - \frac{k}{\hbar} I_{+} +  \bar{g} {\gamma}^5 {\delta(z)}
\right){\psi}(z,t)=0,
\label{onedim}
\end{equation}
$\bar{g} = g/(c{\hbar})$ being the dimensionless coupling constant.

Performing the transformation
\begin{equation}
{\psi}(z,t) \to \tilde{\psi}(z,t)
\equiv \left( I - \frac{{\mu}}{\sqrt{2}} {\gamma}^4 \right) e^{\frac{\rm i}{\hbar} E_0 t} {\psi}(z,t),
\label{transf}
\end{equation}
where ${\mu} = \sqrt{m_0/m}$, and $m_0 = E_0/c^2$ is the rest mass associated with the rest energy, we can bring
Eq.(\ref{onedim}) into the form
\begin{equation}
\left(\frac{\rm i}{c} {\gamma}^4 \frac{\partial}{{\partial}t} +
{\rm i} \gamma^{3} \frac{\partial}{{\partial}z} + \frac{mc}{\hbar} {\gamma}^5 +  \bar{g} {\Gamma}^5 {\delta(z)}
\right)\tilde{\psi}(z,t)=0,
\label{onedimtr}
\end{equation}
which is the LL equation with the rest energy dependent effective potential, with
\[
{\Gamma}^5 \equiv {\gamma}^5 + {\mu} \sqrt{2} I + {\mu}^2 {\gamma}^4.
\]
For a free non-relativistic Fermi field $(g=0)$, the transformation given by Eq.(\ref{transf}) removes
the rest energy term from the GLL equation. However, in the case of pseudoscalar potentials, the rest
energy contributes non-trivially to the external potential term, and this affects the energy spectrum
and the scattering amplitudes.

\section{Bound States}

Representing the field $\tilde{\psi}(z,t)$ as
\[
\tilde{\psi}(z,t)=
\left(
\begin{array}{c}
\tilde{\psi}_{1}(z,t) \\ \tilde{\psi}_{2}(z,t)
\end{array}
\right),
\]
where $\tilde{\psi}_{1}(z,t)$, $\tilde{\psi}_{2}(z,t)$ are $2$-component fields, and introducing their linear combinations
\begin{eqnarray}
\tilde{\eta}_{1}(z,t) & \equiv & \tilde{\psi}_{1}(z,t) + \tilde{\psi}_{2}(z,t) \nonumber \\
\tilde{\eta}_{2}(z,t) & \equiv & ({\mu}+1) \tilde{\psi}_{1}(z,t) + ({\mu}-1) \tilde{\psi}_{2}(z,t) \nonumber
\end{eqnarray}
we can rewrite Eq.(\ref{onedimtr}) in the component form as a system of two equations:
\begin{eqnarray}
\left( {\rm i} \frac{\hbar}{c} \frac{\partial}{{\partial}t} + \frac{\mu}{\sqrt{2}} k \right) \tilde{\eta}_{1}(z,t) +  p_{-} \tilde{\eta}_{2}(z,t) & = & 0, \nonumber \\
p_{+} \tilde{\eta}_{1}(z,t) - \left( mc + \bar{g} {\hbar} {\delta}(z) \right) \tilde{\eta}_{2}(z,t) & = & 0.
\label{system1}
\end{eqnarray}
where
\[
p_{\pm} \equiv \frac{1}{\sqrt{2}} \left( {\hbar} {\sigma}^3 \frac{\partial}{{\partial}z} \pm k \right),
\]
Only one of the components is dynamically independent. Its time evolution determines the time evolution of another component as well.

For stationary states corresponding to energy $E$,
\begin{eqnarray}
\tilde{\eta}_{1}(z,t) & = & \tilde{\eta}_{1}(z) e^{-\frac{\rm i}{\hbar}Et}, \nonumber \\
\tilde{\eta}_{2}(z,t) & = & \tilde{\eta}_{2}(z) e^{-\frac{\rm i}{\hbar}Et}. \nonumber
\end{eqnarray}
Substituting this into Eq.(\ref{system1}) and eliminating $\tilde{\eta}_{1}(z)$ in favor of $\tilde{\eta}_{2}(z)$,
we get
\begin{equation}
\left( - \frac{{\hbar}^2}{2m} \frac{d^2}{dz^2} - {\Lambda} {\delta}(z) \right)
\tilde{\eta}_{2}(z) = E \tilde{\eta}_{2}(z),
\label{schreq}
\end{equation}
i.e. the one-dimensional Schrodinger equation with ${\delta}$-function potential and the energy dependent coupling constant
\begin{equation}
{\Lambda} \equiv \frac{\bar{g} {\hbar}}{mc} (E + E_0).
\label{coupc}
\end{equation}
For bound states, $E<0$, and such bound state exists if the ${\delta}$-function potential is attractive,
${\Lambda}>0$ \cite{flugge}. This gives us two options: $|E|<E_0$, $g>0$ and $|E|>E_0$, $g<0$.
%
%
%
%

%
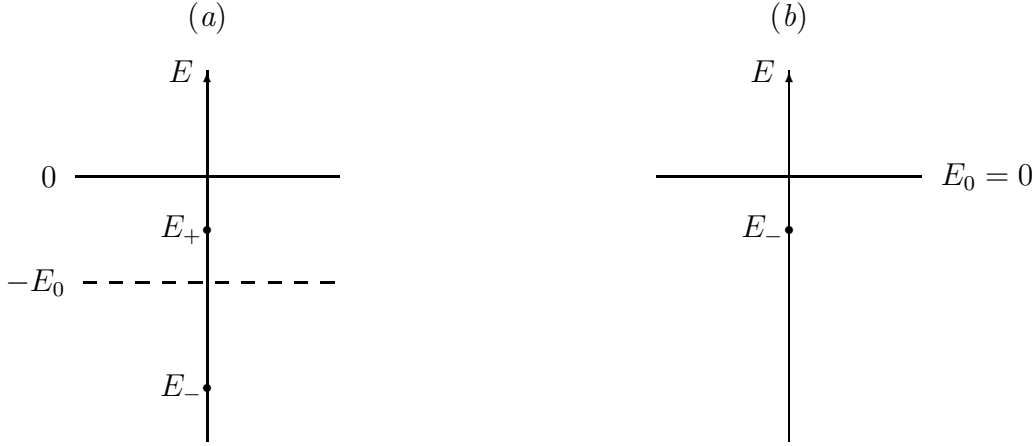
\begin{figure}[hbtp]
\vspace{-3cm}
\begin{picture}(200,400)(10,20)
\put (40,280){\line(1,0){50}}
\put (260,280){\line(1,0){50}}

\put (90,280){\vector(0,1){40}}
\put (90,340){\makebox(0,0){({\it a})}}
\put (80,320){\makebox(0,0){$E$}}
\put (90,280){\line(0,-1){100}}
\put (310,280){\vector(0,1){40}}
\put (385,280){\makebox(0,0){$E_{0}=0$}}
\put (310,340){\makebox(0,0){({\it b})}}
\put (300,320){\makebox(0,0){$E$}}
\put (310,280){\line(0,-1){100}}
\put (90,260){\circle*{3}}
\put (90,200){\circle*{3}}
\put (30,280){\makebox(0,0){$0$}}
\put (80,200){\makebox(0,0){$E_{-}$}}
\put (300,260){\makebox(0,0){$E_{-}$}}
\put (310,260){\circle*{3}}
\put (80,260){\makebox(0,0){$E_{+}$}}
\put (25,240){\makebox(0,0){$-E_{0}$}}

\put (43,240){\line(1,0){5}}
\put (53,240){\line(1,0){5}}
\put (63,240){\line(1,0){5}}
\put (73,240){\line(1,0){5}}
\put (83,240){\line(1,0){5}}
\put (93,240){\line(1,0){5}}
\put (103,240){\line(1,0){5}}
\put (113,240){\line(1,0){5}}
\put (123,240){\line(1,0){5}}
\put (133,240){\line(1,0){5}}


\put (90,280){\line(1,0){50}}
\put (310,280){\line(1,0){50}}

\end{picture}
\vspace{-5.5cm}
\caption{\it Schematic representation of bound states:(a) ${\mu}^2=1$; $E_{+}= - \frac{E_0}{2}$, $\bar{g}=2$;
$E_{-}=-2E_0$, $\bar{g}=-2$,(b) ${\mu}^2=0$; $E_{-}=-\frac{1}{2}mc^2$; $\bar{g}=-2$.}
\end{figure}
%



%
%
%
%

The bound state energy is given by
\begin{equation}
E = - \frac{m{\Lambda}^2}{2{\hbar}^2}.
\end{equation}
Solving it for $E$, we get
\begin{equation}
E_{+} = \frac{E_0}{{\mu}^2 {\bar{g}}^2} \Big( \sqrt{1 + 2{\mu}^2 {\bar{g}}^2} - 1 \Big) - E_0
\qquad {\rm for} \qquad g>0
\label{energyone}
\end{equation}
and
\begin{equation}
E_{-} = - \frac{E_0}{{\mu}^2 {\bar{g}}^2} \Big( \sqrt{1 + 2{\mu}^2 {\bar{g}}^2} + 1 \Big) - E_0
\qquad {\rm for} \qquad g<0,
\label{energytwo}
\end{equation}
so, for non-zero values of ${\mu}^2$, the bound state exists for both signs of the original coupling constant g.
As ${\mu}^2 \to 0$ (or $m_0 \to 0$), $E_{+} \to 0$, and the $g>0$ bound state merges with the continuum
(see Figure 1).




The normalized bound state wavefunctions are
\begin{equation}
\tilde{\eta}_{2,\pm}(z) = {\eta}_{0,\pm} {\left(\frac{2m|E_{\pm}|}{{\hbar}^2}\right)}^{1/4}
\exp\left\{- \frac{\sqrt{2m|E_{\pm}|}}{\hbar} |z|\right\},
\end{equation}
where the subscripts "+" and "-" correspond to the $g>0$ and $g<0$ bound states, respectively (see Figure 2). The column
matrices ${\eta}_{0,\pm}$ obey the condition
\[
{\eta}^{\dagger}_{0,\pm} {\eta}_{0,\pm} = 1
\]
and can be taken as either "spin-up"
$\left(
\begin{array}{c}
1 \\ 0
\end{array}
\right)$
or "spin-down"
$\left(
\begin{array}{c}
0 \\ 1
\end{array}
\right)$
columns.

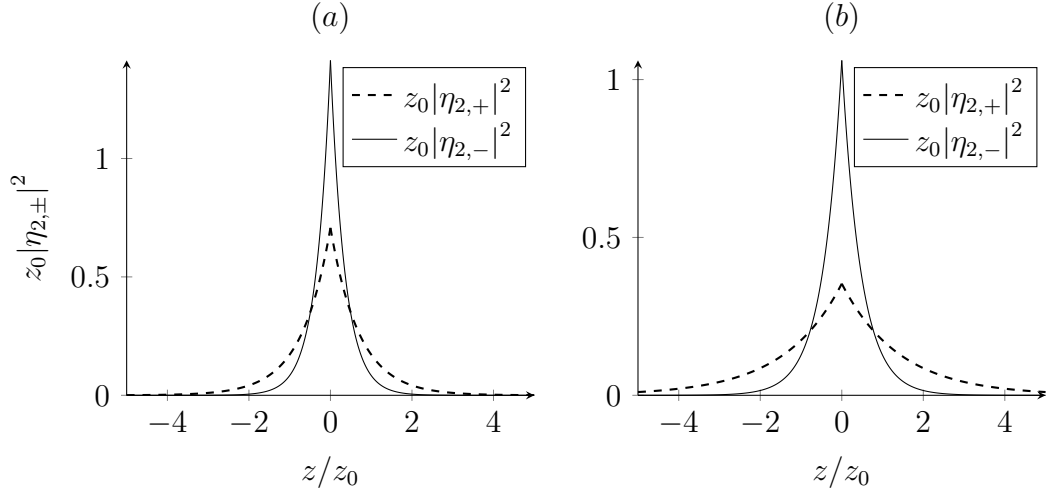
\begin{figure}
\begin{tikzpicture}
\begin{axis}[
axis lines = left,
xlabel = $z/z_{0}$,
ylabel = {$z_{0}{|{\eta}_{2,\pm}|}^2$},
title = {$(a)$},
]
\addplot[thick,dashed][
domain = 0:5,
samples = 200,
]
{sqrt(2)*0.5*exp(-sqrt(2)*x)};
\addlegendentry{$z_{0}{|{\eta}_{2,+}|}^2$}

\addplot[
domain = 0:5,
samples = 200,
]
{sqrt(2)*exp(- 2*sqrt(2)*x)};
\addlegendentry{$z_{0}{|{\eta}_{2,-}|}^2$}

\addplot[thick,dashed][
domain = -5:0,
samples = 200,
]
{sqrt(2)*0.5*exp(sqrt(2)*x)};

\addplot[
domain = -5:0,
samples = 200,
]
{sqrt(2)*exp(2*sqrt(2)*x)};

\end{axis}
\end{tikzpicture}
\hskip 10pt 
\begin{tikzpicture}
\begin{axis}[
axis lines = left,
xlabel = $z/z_{0}$,
title = {$(b)$},
]
\addplot[thick,dashed][
domain = 0:5,
samples = 200,
]
{sqrt(2)*0.5*0.5*exp(-sqrt(2)*0.5*x)};
\addlegendentry{$z_{0}{|{\eta}_{2,+}|}^2$}

\addplot[
domain = 0:5,
samples = 200,
]
{3*sqrt(2)*0.25*exp(- 3*sqrt(2)*0.5*x)};
\addlegendentry{$z_{0}{|{\eta}_{2,-}|}^2$}

\addplot[thick,dashed][
domain = -5:0,
samples = 200,
]
{sqrt(2)*0.25*exp(sqrt(2)*0.5*x)};

\addplot[
domain = -5:0,
samples = 200,
]
{3*sqrt(2)*0.25*exp(3*sqrt(2)*0.5*x)};

\end{axis}
\end{tikzpicture}
\caption{\it The probability density of bound states for $\bar{g}=-2$ (the solid line) and $\bar{g}=2$ (the dashed line): (a) ${\mu}^2 = 1$, (b) ${\mu}^2=\frac{3}{8}$.
The dimensionless variables are used, $z_0 \equiv \frac{\hbar}{mc\sqrt{2}}$.}
\end{figure}

Since Eq.(\ref{schreq}) is covariant under the reflection $z \to -z$, the component $\tilde{\eta}_{2,\pm}(z)$ has
positive parity. The dependent component
\[
\tilde{\eta}_{1,\pm}(z) = \frac{c}{\sqrt{2}(E_{\pm} + E_0)}
\left( k + \sqrt{2m|E_{\pm}|} {\epsilon}(z) {\sigma}^3 \right) \tilde{\eta}_{2,\pm}(z),
\]
where ${\epsilon}(z) = {\theta}(z) - {\theta}(-z)$, and ${\theta}(z)$ is the Heaviside step function, is not continuous at $z=0$. It cannot be characterized by a definite parity. This is the result of rest energy contribution.
In the limit ${\mu}^2 \to 0$, when the rest energy effects disappear, $\tilde{\eta}_{1,-}(z)$ has the
parity opposite to the parity of $\tilde{\eta}_{2,-}(z)$.

\section{Scattering States}

For scattering states, $E>0$. Assuming that the Fermi particles are incident from the left, we represent the solution of Eq.(\ref{schreq}) as
\begin{equation}
\tilde{\eta}_{2}(z) = {\eta}_0
\left\{
\begin{array}{cc}
e^{\frac{\rm i}{\hbar} pz} + R e^{-\frac{\rm i}{\hbar} pz} & {\rm for} \hspace{5 mm} z<0,\\
S e^{\frac{\rm i}{\hbar} pz} & {\rm for} \hspace{5 mm} z>0,
\end{array}
\label{scatsol}
\right.
\end{equation}
where $p \equiv \sqrt{2mE}$, and $R$ and $S$ are the amplitudes of the reflected and transmitted waves, respectively. As before, ${\eta}_0$ specifies the orientation of spin of incoming particles, and ${\eta}^{\dagger}_0 {\eta}_0 = 1$.

The condition of continuity of $\tilde{\eta}_2(z)$ at $z=0$ gives
\begin{equation}
1 + R = S.
\label{firstcond}
\end{equation}
The derivative $\tilde{\eta}_2^{\prime} \equiv d{\tilde{\eta}}_2/dz$ is discontinuous at $z=0$. Substituting the ansatz (\ref{scatsol}) into Eq.(\ref{schreq}), integrating the equation from $-{\varepsilon}$ to ${\varepsilon}$ and taking the limit ${\varepsilon} \to 0^{+}$, this yields
\[
\tilde{\eta}_2^{\prime}(0^{+}) - \tilde{\eta}_2^{\prime}(0^{-}) = - \frac{2m{\Lambda}}{{\hbar}^2} \tilde{\eta}_2(0)
\]
\begin{figure}[hbtp]
\vspace{-3cm}
\begin{picture}(200,400)(10,20)
\put (20,280){\line(1,0){140}}
\put (240,280){\line(1,0){140}}

\put (90,280){\vector(0,1){40}}
\put (90,340){\makebox(0,0){({\it a})}}
\put (90,280){\line(0,-1){40}}
\put (310,280){\vector(0,1){40}}
\put (310,340){\makebox(0,0){({\it b})}}
\put (310,280){\line(0,-1){40}}
\put (90,280){\circle*{3}}
\put (310,280){\circle*{3}}
\put (150,310){\makebox(0,0){$E$}}
\put (370,310){\makebox(0,0){$E$}}
\put (150,310){\circle{14}}
\put (370,310){\circle{14}}
\put (95,272){\makebox(0,0){$0$}}
\put (315,272){\makebox(0,0){$0$}}
\put (30,280){\circle*{3}}
\put (30,270){\makebox(0,0){$E_{-}$}}
\put (70,280){\circle*{3}}
\put (280,280){\circle*{3}}
\put (280,270){\makebox(0,0){$E_{+}$}}
\put (70,270){\makebox(0,0){$-\frac{2mc^2}{{\bar{g}}^2}$}}

\put (33,275){\line(1,0){5}}
\put (43,275){\line(1,0){5}}
\put (53,275){\vector(1,0){5}}

\put (283,275){\line(1,0){5}}
\put (293,275){\line(1,0){5}}
\put (303,275){\vector(1,0){5}}

\thicklines
\put (90,280){\vector(1,0){70}}
\put (310,280){\vector(1,0){70}}

\end{picture}
\vspace{-7.5cm}
\caption{\it The physical sheet of the Riemann surface of the analytic functions $R(E)$ and $S(E)$: (a) $g<0$, (b) $g>0$. The exact location of the poles depends on the value of ${\mu}^2$. Both poles move to the right with ${\mu}^2$ decreasing. In the limit ${\mu}^2 \to 0$, $E_{-}$ stops at $-2mc^2/{\bar{g}}^2$, while $E_{+}$ reaches the branch point $E=0$ and disappears.}
\end{figure}
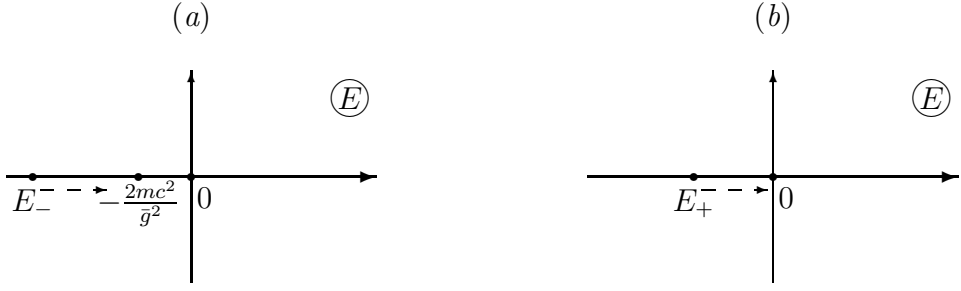
or
\begin{equation}
{\rm i} p (S - 1 + R) = - \frac{2m{\Lambda}}{\hbar} S.
\label{secondcond}
\end{equation}
Solving equations (\ref{firstcond}) and (\ref{secondcond}), we get
\begin{equation}
R = - \frac{m{\Lambda}}{m{\Lambda} + {\rm i}p{\hbar}}, \qquad
S = \frac{{\rm i}p{\hbar}}{m{\Lambda} + {\rm i}p{\hbar}}.
\label{rs}
\end{equation}

Extending both amplitudes to the complex domain, we can consider them as analytic functions of the complex variable $E$:
\begin{eqnarray}
R(E) & = & - \frac{\bar{g} {\hbar}}{2mc} \frac{E+E_0}{\left(\sqrt{E}-{\rm i}{\epsilon}(g) \sqrt{|E_{+}|}\right)
\left(\sqrt{E}+{\rm i}{\epsilon}(g) \sqrt{|E_{-}|}\right)}, \nonumber \\
S(E) & = & \frac{{\rm i}{\hbar}}{\sqrt{2m}} \frac{\sqrt{E}}{\left(\sqrt{E}-{\rm i}{\epsilon}(g) \sqrt{|E_{+}|}\right)
\left(\sqrt{E}+{\rm i}{\epsilon}(g) \sqrt{|E_{-}|}\right)}, \nonumber
\end{eqnarray}
where $E_{+}$ and $E_{-}$ are given by Eqs.(\ref{energyone}) and (\ref{energytwo}). The functions $R(E)$ and $S(E)$ are multi-valued, $E=0$ and $E={\infty}$ being the branch points. If we cut the complex $E$-plane along the right half of the real axis $E>0$, then on the physical sheet for the points on the upper edge of the cut the phase of E is equal to zero. In these points, the values of $R(E)$ and $S(E)$ coincide with those of the physical reflection and transmission amplitudes.

Both functions have two poles determined by the conditions
\[
\sqrt{E} = {\rm i} \epsilon(g) \sqrt{|E_{+}|}
\]
and
\[
\sqrt{E} = - {\rm i} \epsilon(g) \sqrt{|E_{-}|}
\]
The pole $E=-|E_{+}|$ is located on the physical sheet for $g>0$ (the phase is $\pi$) and on the non-physical one for $g<0$ (the phase is $3{\pi}$). The situation with the pole $E=-|E_{-}|$ is opposite: it is on the physical sheet for $g<0$, and it moves to the non-physical sheet, i.e. becomes virtual, for $g>0$ (see Figure 3).

The coefficients of reflection and transmission are
\[
|R(E)|^2 = \frac{(E + E_0)^2}{(E + |E_{+}|)(E + |E_{-}|)}
\]
and
\[
|S(E)|^2 = \frac{2mc^2}{{\bar{g}}^2} \frac{E}{(E + |E_{+}|)(E + |E_{-}|)}.
\]
In the limit ${\mu}^2 \to 0$, they take the form
\begin{eqnarray}
|R(E)|^2 & = & \frac{E}{E + \frac{2mc^2}{{\bar{g}}^2}}, \nonumber \\
|S(E)|^2 & = & \frac{2mc^2}{{\bar{g}}^2} \frac{1}{E + \frac{2mc^2}{{\bar{g}}^2}}. \nonumber
\end{eqnarray}

As $E \to 0$, $|R(E)|^2$ approaches $1$ for non-zero values of ${\mu}^2$ and $0$ for ${\mu}^2 = 0$. The reason of such behavior is in the energy dependence of the effective coupling constant ${\Lambda}$. Without the rest energy contribution, ${\Lambda}$ vanishes as $E \to 0$. The ${\delta}$-function potential disappears, and this results in complete transmission. With the rest energy taken into account, there is a non-zero coupling for $E \to 0$ as well, and it yields complete reflection. This is in agreement with the theorem on low momentum scattering in the one-dimensional Schr\"odinger equation by an even potential well \cite{senn}.

As $E \to \infty$, the magnitude of ${\Lambda}$ becomes infinitely large, and $|R(E)|^2 \to 1$ in both cases, ${\mu}^2 \neq 0$ and ${\mu}^2 = 0$. This is valid for both signs of the original coupling constant $g$.

\section{Discussion}

1. The one-dimensional GLL equation with pseudoscalar ${\delta}$-function potential is reduced to the Schr\"odinger
equation with an effective ${\delta}$-function potential the coupling constant of which depends on both the energy of the Fermi particle and its rest energy. The energy dependence of the effective potential appears here in the same way as in the case of the Dirac equation reduced to the Pauli-Schr\"odinger one \cite{pauli},\cite{bethe}.

For the vanishing rest energy $E_0=0$, the transition from the negative $(g<0)$ to positive $(g>0)$ value of the original coupling constant reverses the sign of the effective coupling constant as well. Being attractive for
$g<0, E<0$, and able to bind the Fermi particle, the effective potential becomes repulsive for $g>0, E<0$, and fails to confine it. There is a single bound state in this case, and this is similar to the case of one-dimensional Schr\"odinger equation with ${\delta}$-function potential.

For non-zero values of rest energy, the picture changes. The value of energy at which the effective coupling constant vanishes is shifted from $E=0$ to $E=-E_0$. This divides the negative energies into two intervals:
$E<-E_0$ and $-E_0<E<0$. For energies below $(-E_0)$, the effective potential is attractive for $g<0$ and repulsive for $g>0$. For energies above $(-E_0)$, the situation is opposite, and the effective potential becomes attractive for $g>0$. This provides us with two binding energies, one below $(-E_0)$ and another one above $(-E_0)$. In the limit $E_0 \to 0$, the bound state below $(-E_0)$ reduces to the one we had before in the case of vanishing rest energy, while the bound state above $(-E_0)$ disappears. The existence of an additional bound state for non-zero values of rest energy agrees with the statements done in \cite{sar}.

2. For scattering states, if the energy of the Fermi particle is infinitely large $(E \to \infty)$, the effective
${\delta}$-function potential becomes very opaque $({\Lambda} \to \infty)$, and we approach complete reflection. This is opposite to the case of one-dimensional Schr\"odinger equation with ${\delta}$-function potential where complete transmission is observed for $E \to \infty$ and a finite value of the coupling constant. On the contrary, if the energy of the Fermi particle is infinitely small $(E \to 0)$ and its rest energy is non-zero, we approach complete reflection for both equations.

\end{document}